\documentstyle[12pt]{article}
\begin{document}\bibliographystyle{plain}\begin{titlepage}
\renewcommand{\thefootnote}{\fnsymbol{footnote}}\hfill\begin{tabular}{l}
HEPHY-PUB 761/02\\UWThPh-2002-38\\December 2002\end{tabular}\\[3cm]\Large
\begin{center}{\bf INFLUENCE OF THE SOMMERFELD CORRECTIONS TO THE ELECTRIC
POLARIZABILITY OF THE PION}\\\vspace{2cm}\large{\bf Wolfgang
LUCHA\footnote[1]{\normalsize\ {\em E-mail address\/}:
wolfgang.lucha@oeaw.ac.at}}\\[.3cm]\normalsize Institut f\"ur
Hochenergiephysik,\\\"Osterreichische Akademie der
Wissenschaften,\\Nikolsdorfergasse 18, A-1050 Wien, Austria\\[1cm]\large{\bf
Franz F.~SCH\"OBERL\footnote[2]{\normalsize\ {\em E-mail address\/}:
franz.schoeberl@univie.ac.at}}\\[.3cm]\normalsize Institut f\"ur Theoretische
Physik, Universit\"at Wien,\\Boltzmanngasse 5, A-1090 Wien, Austria\vfill
\begin{quote}{\it To appear in the Proceedings of the International
Conference on ``Quark Confinement and the Hadron Spectrum V,'' September 10
-- 14, 2002, Gargnano, Italy}\end{quote}

\vspace{1cm}

{\normalsize\bf Abstract}\end{center}\small The valence-quark contribution to
the electric polarizability of the charged pion~in a semirelativistic
description is shown to be smaller than its nonrelativistic
limit.\vspace{3ex}

\renewcommand{\thefootnote}{\arabic{footnote}}\end{titlepage}

\normalsize

\section{Electric Polarizability: Probe of the Pion's Internal
Structure}Exposing a physical system composed of electrically charged
constituents~to an external electric field $\mbox{\boldmath{$E$}}$ induces
the electric dipole moment\footnote{Using the {\it Heaviside--Lorentz\/}
system of units for electromagnetic quantities, the unit of electric charge,
$e,$ is related to the electromagnetic fine structure constant $\alpha$
by$$\alpha\equiv e^2/4\pi\simeq 1/137\ .$$}
$$\mbox{\boldmath{$d$}}=4\pi\,\kappa\,\mbox{\boldmath{$E$}}\ ;$$the constant
of proportionality herein, the electric polarizability $\kappa,$ measures the
system's rigidity under deformation and represents a clue to the internal
structure of composite systems. Chiral perturbation theory, the low-energy
limit of quantum chromodynamics, makes a precise unambiguous prediction for
the electric polarizability $\kappa_{\pi^\pm}$ of the charged pion $\pi^\pm.$
Consequently, $\kappa_{\pi^\pm}$ forms a crucial test of quantum
chromodynamics.\footnote{The implications of chiral symmetry for the pion's
electromagnetic polarizabilities~are reviewed in Ref.~\cite{Holstein}.}
Experiment studies $\kappa_{\pi^\pm}$ by reactions based on Compton
scattering$$\gamma+\pi^\pm\to\gamma+\pi^\pm$$or crossed processes. Our
weighted average of the few existing experimental measurements of the
electric polarizability of the charged pion reads \cite{Lucha:RelPiPol}
$$\kappa_{\pi^\pm}^{\rm exp}=(4.3\pm1.2)\times10^{-4}\;\mbox{fm}^3\ .$$The
mean value of the theoretical predictions of all {\it
quantum-field-theoretic\/} descriptions of the mesons, for the electric
polarizability of the charged pion,$$\kappa_{\pi^\pm}^{\rm QFT}=5.5\times
2.3^{\pm1}\times 10^{-4}\;\mbox{fm}^3\ ,$$is in reasonable agreement with
experiment \cite{Lucha:RelPiPol}.

\section{Electric Polarizabilities of Mesons within Quark Potential
Models}Quark potential models of mesons are, in principle, only capable to
describe the contributions of the valence quarks to the meson's electric
polarizability. In a naive static {\it nonrelativistic\/} potential model
this quark-core contribution accounts for only a fraction 1/80 of the
charged-pion electric polarizability \cite{Leeb}.

We examine the effects of kinematics by relaxing the tight nonrelativistic
bound towards more reliable {\it semirelativistic\/} treatments of mesons
viewed~as constituent quark-antiquark bound states, with a Hamiltonian
$$H=H_0+W\ ,$$where the bound state is described by the unperturbed
Hamiltonian $H_0,$~its interaction with the external electric field
$\mbox{\boldmath{$E$}}$ is encoded in a perturbation~$W.$ For equal-mass
bound-state constituents with electric charges $q_1,$ $q_2$ located at
$\mbox{\boldmath{$r$}}_1,$ $\mbox{\boldmath{$r$}}_2,$ this residual dipole
interaction reads$$W=\mbox{$\frac{1}{2}$}\,(q_2-q_1)\,e\,
(\mbox{\boldmath{$r$}}_1-\mbox{\boldmath{$r$}}_2)\cdot\mbox{\boldmath{$E$}}\
.$$The semirelativistic Hamiltonian $H_0$ governing the dynamics of two
quarks with mass $m,$ relative momentum $\mbox{\boldmath{$p$}},$ and relative
distance $r$ reads, in their center-of-momentum
frame,$$H_0=2\,\sqrt{\mbox{\boldmath{$p$}}^2+m^2}+V(r)\ .$$The strong
interactions between the quarks are described by a static central potential
$V(r).$ We analyze the change of the quark-model prediction for~the
charged-pion electric polarizability brought about when taking into account
the Sommerfeld corrections$$\sqrt{\mbox{\boldmath{$p$}}^2+m^2}-m-
\frac{\mbox{\boldmath{$p$}}^2}{2\,m}$$to the kinetic energy.

The energy levels of the full Hamiltonian $H$ are calculated by expanding
about the (normalized) eigenstates $|\phi\rangle$ of the unperturbed
Hamiltonian $H_0.$ For spherically symmetric states---such as the ground
state---the interaction with the external field $\mbox{\boldmath{$E$}}$ then
entails, in lowest order in $W,$ the energy~shift \cite{Lucha:RelPiPol}
$$\Delta E=-\frac{1}{2}\,\mbox{\boldmath{$d$}}\cdot\mbox{\boldmath{$E$}}
=-\frac{\langle\phi|W^2|\phi\rangle^2} {\langle\phi|[W,H_0]\,W|\phi\rangle}\
.$$Assuming, w.l.o.g., $\mbox{\boldmath{$E$}}$ to be parallel to the $z$-axis
of the employed coordinate frame allows to read off the electric
polarizability $\kappa$ of a meson consisting~of equal-mass quarks from the
contribution of order $\mbox{\boldmath{$E$}}^2$ to $\Delta E$ (for
details,~and the generalization to the case of quarks of unequal masses,
consult Ref.~\cite{Lucha:RelPiPol}):$$\kappa=\frac{\alpha}{18}\,(q_2-q_1)^2\,
\frac{\langle\phi|r^2|\phi\rangle^2} {\langle\phi|[z,H_0]\,z|\phi\rangle}\ .$$

Assuming the ground state of the meson under study to be described~by a real
wave function $\phi(p),$ $p\equiv|\mbox{\boldmath{$p$}}|,$ the electric
polarizability $\kappa$ of this meson is thus determined by the magnitude of
the expectation values $\langle\phi|r^2|\phi\rangle$ and
\cite{Lucha:RelPiPol}
\begin{eqnarray*}\langle\phi|[z,H_0]\,z|\phi\rangle&=&\frac{1}{3}\int{\rm
d}^3p\,\frac{2\,p^2+3\,m^2}{\left(p^2+m^2\right)^{3/2}}\,\phi^2(p)\\[1ex]&\le&
\langle\phi|[z,H_0]\,z|\phi\rangle_{\rm NR}=\frac{1}{m}\ .\end{eqnarray*}As
indicated, the latter is bounded from above by its nonrelativistic limit.

\section{Representative Results, Brief Discussion and Conclusion}The
lowest-energy eigenfunction $\phi(p)$ of the semirelativistic
Hamiltonian~$H_0$ is obtained variationally with a 25-dimensional trial space
spanned by basis functions involving generalized Laguerre polynomials
\cite{Lucha:LagTS}. Table~\ref{Tab:Pots} compares for some typical interquark
potentials $V(r)$ the electric polarizability $\kappa_{\pi^\pm}$~of the
charged pion computed from relativistic and nonrelativistic kinematics.

\begin{table}[h]\caption{Ratio of the electric polarizability of the charged
pion $\pi^\pm$ in various nonrelativistic ($\kappa_{\rm NR}$) and
relativistic ($\kappa$) potential models, for the canonical light-quark
constituent mass $m=0.336\;\mbox{GeV}$ and ``reasonable'' interquark
potential parameters (cf.~Ref.~\cite{Lucha:QBS}).}\label{Tab:Pots}
\begin{center}\begin{tabular}{cllll}\hline\hline&&&&\\[-1.5ex]
\multicolumn{1}{c}{potential}&\multicolumn{1}{l}{harmonic oscillator}&
\multicolumn{1}{l}{Coulomb}&\multicolumn{1}{l}{linear}&
\multicolumn{1}{l}{funnel}\\[1ex]\hline&&&&\\[-1.5ex]
$V(r)$&$0.5\,r^2$&$-\displaystyle\frac{0.413}{r}$&$0.15\,r$&
$-\displaystyle\frac{0.413}{r}+0.15\,r$\\[2ex]
$\displaystyle\frac{\kappa_{\rm NR}}{\kappa}$ &1.02&1.08&1.2&1.36\\[2ex]
$|\delta|$&$3\times 10^{-3}$&$2\times 10^{-4}$&$6\times 10^{-10}$&$6\times
10^{-5}$\\[1ex]\hline\hline\end{tabular}\end{center}\end{table}

The accuracy of approximate eigenstates $|\varphi\rangle$ of Hamiltonians
$$H=T+V$$involving kinetic terms $T(\mbox{\boldmath{$p$}})$ and potentials
$V(\mbox{\boldmath{$x$}})$ is quantified, according~to a criterion based on
the relativistic virial theorem \cite{Lucha:RVT}, by the nonzero value~of
\cite{Lucha:Accuracy}$$\delta\equiv1-\frac{
\left\langle\varphi\left|\mbox{\boldmath{$x$}}\cdot\frac{\partial}
{\partial\mbox{\boldmath{$x$}}}V(\mbox{\boldmath{$x$}})
\right|\varphi\right\rangle}{
\left\langle\varphi\left|\mbox{\boldmath{$p$}}\cdot\frac{\partial}
{\partial\mbox{\boldmath{$p$}}}T(\mbox{\boldmath{$p$}})
\right|\varphi\right\rangle}\ .$$The very tiny values of $|\delta|$ in
Table~\ref{Tab:Pots} give us great confidence in our results.

From Table~\ref{Tab:Pots}, we conclude that the Sommerfeld corrections {\em
diminish\/} the quark-core contribution to the pion electric polarizability.
Similar findings \cite{Lucha:RelPiPol} are expected when including also
relativistic corrections to the interaction \cite{Lucha91RelTreat}.

\end{document}